\documentclass[aps,prl,twocolumn,showpacs,floatfix,superscriptaddress]{revtex4-1}

\usepackage[dvips]{graphics}
\usepackage{epsfig}
\usepackage{times}
\usepackage{amsmath,amssymb}
\usepackage{amsthm}
\usepackage{mathrsfs}
\usepackage{multirow}

\newcommand{\bfk}{{\mathbf{k}}}

\newcommand{\varH}{{\mathcal{H}}}

\newcommand{\ket}[1]{\left|#1\right\rangle}
\newcommand{\Braket}[1]{\mathinner{\langle{\textstyle#1}\rangle}}

\newcommand{\figref}[1]{Fig.~\ref{#1}}
\newcommand{\figsref}[1]{Figs.~\ref{#1}}

\let\up\uparrow
\let\down\downarrow

\begin{document}

\title{Kondo Effect in a Quantum-Dot-Topological-Superconductor Junction}

\author{Minchul Lee}
\affiliation{Department of Applied Physics, College of Applied Science, Kyung Hee University, Yongin 446-701, Korea}
\author{Jong Soo Lim}
\affiliation{Institut de F\'{i}sica Interdisciplinar i de Sistemes Complexos IFISC (CSIC-UIB), E-07122 Palma de Mallorca, Spain}
\author{Heunghwan Khim}
\affiliation{Department of Physics, Korea University, Seoul 136-701, Korea}
\author{Rosa L\'opez}
\affiliation{Institut de F\'{i}sica Interdisciplinar i de Sistemes Complexos IFISC (CSIC-UIB), E-07122 Palma de Mallorca, Spain}
\affiliation{Departament de F\'{i}sica, Universitat de les Illes Balears, E-07122 Palma de Mallorca, Spain}

\date{\today}

\begin{abstract}
  We investigate the dynamical and transport features of a Kondo dot
  side-coupled to a topological superconductor (TS). The Majorana fermion
  states (MFS) formed at the ends of the TS are found to be able to alter the
  Kondo physics profoundly:
  For the ideal setup where the MFS do not overlap ($\epsilon_m=0$) a finite
  dot-MFS coupling $(\Gamma_m)$ reduces the unitary-limit value of the linear
  conductance by exactly a factor 3/4 in the Kondo-dominant regime
  $(\Gamma_m<T_K)$, where $T_K$ is the Kondo temperature. In the
  Majorana-fermion dominant phase ($\Gamma_m>T_K)$, on the other hand, the
  spin-split Kondo resonance takes place due to the MFS-induced Zeeman
  splitting, which is a genuine many-body effect of the strong Coulomb
  interaction and the topological superconductivity. We find that the original
  Kondo resonance is fully restored once the MFSs are strongly hybridized
  ($\epsilon_m > \Gamma_m$). This unusual interaction between the Kondo effect
  and the MFS can thus serve to detect the Majorana fermions unambiguously and
  quantify the degree of overlap between the MFSs in the TS.
\end{abstract}

\maketitle

\paragraph{Introduction.---}
One of the most paradigmatic effects in condensed matter physics is the
celebrated \textit{Kondo effect}. The ground state of a metal that contains
magnetic impurities consists of a many-body singlet state where the localized
impurities are entangled with the conducting states \cite{Hewson:1993}.  The
Kondo effect has been observed in manufactured nanostructures such as quantum
dots (QDs) \cite{Kouwenhoven:1997,GoldhaberGordon:1998a,GoldhaberGordon:1998b,
  Cronenwett:1998,Schmid:1998,vanderWiel:2000}, carbon nanotubes
\cite{Nygard:2000,Odom:2000,JarilloHerrero:2005}, nanowires
\cite{Klochan:2011}, and so on. The great advantage of the observation of the
Kondo effect in artificial set-ups is its high tunability and control by means
of the application of electrical gates or external fields (e.g., magnetic
field, an ac signal) that drives the Kondo state towards nonequilibrium
situations \cite{Ralph:1994,Kogan:2004,DeFranceschi:2002,Sanchez:2005}.
Nevertheless, the Kondo physics can be modified not only externally but also
intrinsically by utilizing different type of contacts materials, say,
superconducting or ferromagnetic ones
\cite{Park:2002,Choi:2004,Martinek:2003,Zitko:2012,Lim:2011,Zitko:2010a}.

Since the discovery of the topological materials
\cite{Alicea:2010a,Hasan:2010} there has been a
large number of proposals for the physical realization of low-energy
quasiparticle excitations behaving as Majorana fermions
\cite{Kitaev:2001,Fu:2008,Beenakker:2012,Alicea:2012}.  Recently, Mourik \textit{et. al.}
\cite{Mourik:2012} reported the detection of such quasiparticles in InSb
nanowires brought into proximity with a $s$-wave superconductor in the presence
of both magnetic field and spin-orbit interaction
\cite{Lutchyn:2010,Oreg:2010,Alicea:2010b}. Later on, other groups have also
showed signatures of Majorana physics in similar set-ups
\cite{Deng:2012,Das:2012,Rokhinson:2012}.  Therefore, it is quite natural to
think about more intricate scenarios by combining Kondo-like artificial
impurities with localized Majorana fermions hosted in topological materials.

\begin{figure}[!t]
  \centering
  \includegraphics[width=4.3cm]{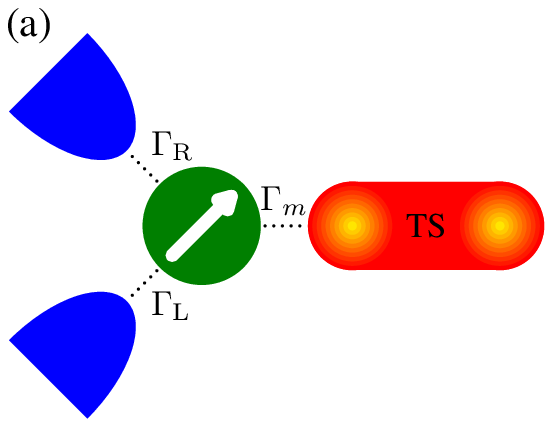}\hspace{.5cm}%
  \includegraphics[width=2.7cm]{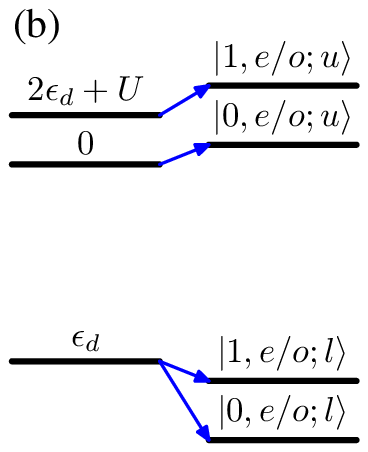}%
  \caption{(Color online) (a) Quantum dot system coupled to two normal-metal
    leads (by tunneling rates $\Gamma_{\rm L(R)}$) and to one end of a floating
    topological superconductor (by QD-MFS tunneling rate $\Gamma_m$).  We
    assume that only the spin-$\down$ component is coupled to the MFS. (b)
    Energy diagram for the (isolated) QD-MFS system showing the induced Zeeman
    field in the dot by the MFS.}
  \label{fig:1}
\end{figure}

The purpose of this work is to analyze a prototypical system to probe the
interplay between Kondo and Majorana physics by means of usual transport
measurements. The system consists of a quantum dot attached to two normal
contacts, in which the Kondo correlations can be developed
\cite{Zazunov:2011,Zitko:2010b,Zitko:2011}, and a topological superconductor
(TS) \cite{Flensberg:2010,Fu:2010,Leijnse:2011,Liu:2011} whose one end is
connected to the QD by a tunneling junction. The TS is floating in a sense that
it is capacitively connected to a gate and no dc current flows through it.  The
two normal leads are to serve as probe of the interplay between the Kondo and
Majorana physics.
In our study not only the deep Kondo regime (previously addressed by an
effective noninteracting theory with limited validity \cite{Golub:2011}) but
also the intermediate regimes where Kondo and Majorana physics are comparable
are thoroughtly examined.  Our findings indicate that the Kondo physics is
dramatically altered depending on the relative strength of \textit{(i)} the
overlap of Majorana fermions states (MFS) ($\epsilon_m$), \textit{(ii)} the
dot-MFS coupling ($\Gamma_m$), and \textit{(iii)} the Kondo temperature
($T_K$). In the ideal Majorana condition $(\epsilon_m=0)$ the half-fermionic
anti-Fano resonance due to the MFS reduces the Kondo value of the linear
conductance by exactly a factor 3/4 in the Kondo-dominant regime
$(\Gamma_m<T_K)$. In the Majorana-fermion dominant phase ($\Gamma_m>T_K)$, on
the other hand, the spin-split Kondo resonance takes place due to the
MFS-induced Zeeman splitting, which is a genuine many-body effect of the strong
Coulomb interaction and the topological superconductivity. We find that once
the MFSs are strongly hybridized ($\epsilon_m > \Gamma_m$) the Kondo effect is
unaffected by the MFS as demonstrated below.

\paragraph{Model.---}
Our system is mapped onto a modified two-fold degenerate Anderson model where
the QD state is coupled to two normal-metal contacts and a topological
superconducting wire that hosts a pair of MFSs at its ends (see
\figref{fig:1}). Due to the helical property of such end-states only one of the
dot spin orientations (say spin-$\down$) hybridizes with the fully
spin-polarized nearest MFS \cite{Flensberg:2010,simon2012}. The Hamiltonian then reads
\begin{align}
  \nonumber
  \varH
  & =
  \sum_{\ell\bfk\mu} \epsilon_\bfk c_{\ell\bfk\mu}^\dag c_{\ell\bfk\mu}
  + \sum_\mu \epsilon_d d_\mu^\dag d_\mu + U n_\up n_\down
  + 2i\epsilon_m \gamma_1\gamma_2
  \\
  & \mbox{}
  + t_m \left(d_\down^\dag\gamma_1+\gamma_1d_\down\right)
  +
  \sum_{\ell\bfk\mu}
  \left(t_\ell d_\mu^\dag c_{\ell\bfk\mu} + (h.c.)\right),
\end{align}
where $c_{\ell\bfk\mu}^\dag$ creates an electron with momentum $\bfk$, energy
$\epsilon_\bfk$, and spin $\mu$ in the $\ell=\rm L,R$ reservoir. Two normal
contacts share a same flat-band structure with a half bandwidth $D$ and density
of states $\rho$. The operator $d_\mu^\dag$ creates an electron with spin $\mu$
in dot, and $n_\mu=d_\mu^\dag d_\mu$ is the dot occupation for spin $\mu$. We
focus on the case of single orbital level with energy $\epsilon_d$ and strong
Coulomb interaction denoted as $U$. The superconducting wire, assumed to be in
the topological state, has two MFSs, $\gamma_1$ and $\gamma_2$ at its two ends:
the MF operators follow the Clifford algebra
$\{\gamma_i,\gamma_j\}=\delta_{ij}$, where $\gamma_i=\gamma_i^\dag$. In terms
of ordinary fermionic operator $f$, they can be written as
$\gamma_1=(f+f^\dag)/\sqrt{2}$, and $\gamma_2=(f-f^\dag)/i\sqrt{2}$.  In
finite-length wires, the two MFSs have a finite overlap between their
wavefunctions so that their coupling can lead a finite gap represented by
$\epsilon_m$.  The dot electron hybridizes \textit{(i)} with the conduction
electrons in the contacts with a tunneling amplitude $t_\ell$, and
\textit{(ii)} with the nearest MFS with a tunneling amplitude $t_m$. Both
couplings define the two tunneling rates: $\Gamma_\ell = \pi \rho t_\ell^2$,
and $\Gamma_m = \pi \rho_{\rm dot} t_m^2$. Throughout our study, we focus on
the Kondo regime, $\epsilon_d < \epsilon_F = 0 < \epsilon_d + U$ and
$\Gamma\equiv\Gamma_{\rm L}+\Gamma_{\rm R} \ll |\epsilon_d|, \epsilon_d+U$ at
zero temperature, where $\epsilon_F$ is the Fermi energy.

For a non-perturbative study of the many-body effect, we adopt the well-known
numerical renormalization group (NRG) method
\cite{Wilson:1975,Krishnamurthy:1980,Hofstetter:2000}: one can refer
Ref.~\onlinecite{Bulla:2008} for a review.
For better efficiency, we exploit the symmetries that our system has:
$[Q_\up,\varH] = [P_\down,\varH] = 0$ where $Q_\up$ and $P_\down$ are charge
number operator for spin-$\up$ electrons and parity operator for the sum
$Q_\down$ of spin-$\down$ electrons and $f$ electrons, respectively. Note that
the QD-MFS hopping changes $Q_\down$ by even numbers only.
For the analysis, we calculate the spectral densities $A_{\mu(m)}= \sum_n
|\Braket{n|d^\dag_\mu (f^\dag)|0}|^2 \delta(\omega{-}E_n{+}E_0)$ and
$A_{zz}=\sum_n|\Braket{n|S_z|0}|^2 \delta(\omega{-}E_n{+}E_0)$, where $\ket{n}$
is the many-body eigenstate with energy $E_n$ and $\ket{0}$ is the ground
state. From the spin-resolved spectral densities the transmission through the
dot can be obtained, $T(\omega)=2\pi\Gamma_{\rm L}\Gamma_{\rm R}/(\Gamma_{\rm
  L}+\Gamma_{\rm R})\sum_\mu A_\mu(\omega)$, and the linear conductance is $G =
(2e^2/h) T(\omega=0)$.

\paragraph{Noninteracting Case ($U=0$).---}
Before addressing the full system we examine simpler cases. In the
noninteracting case, the effect of the MFS on the transport can be handled by
using the spinless model since the MFS is coupled to spin-$\down$ electrons
only \cite{Liu:2011}. Therefore, $A_\up(\omega)$, forming a resonance peak of
width $\Gamma$, is not affected by the QD-MFS coupling.
On the other hand, $A_\down(\omega)$ features the destructive interference
between spin-$\down$ electrons and MFS. In the on-resonant case
($\epsilon_d=0$) and for $\epsilon_m=0$ and small $t_m$, the Fano-like
anti-resonance leads to a half-dip at $\omega=0$: $\pi\Gamma A_\down(0)$ is
reduced from 1 to $1/2$. The dip width is comparable with the resonance width
$\Gamma_m$ in $A_m(\omega)$ which is numerically found to be $\Gamma_m \approx
\pi \rho_{\rm dot} t_m^2$ with $\rho_{\rm dot}=1/\Gamma$.
As $t_m$ increases further ($\Gamma_m>\Gamma$), $A_\down(\omega)$ develops two
side peaks at $\omega\sim\pm\sqrt2 t_m$ which come from the hybridization
between spin-$\down$ dot electron and MFS, while $\pi\Gamma A_\down(0)=1/2$ is
maintained due to the zero-energy MFS \cite{Liu:2011}.
Similarly, $A_m(\omega)$ also exhibits three-peak structure at
$\omega = 0$ and $\pm\sqrt2 t_m$, while the height $\pi\Gamma_m A_m(0)$
increases with $t_m$.
Consequently, the linear conductance $G$ at zero temperature is quantized to
$e^2/h + e^2/2h = 3e^2/2h$ as long as $\epsilon_m = 0$, which signals the
presence of MFS in noninteracting side-coupled nanowire setups.




\paragraph{Isolated QD-MFS System ($t_\ell=0$).---}
Decoupled from the leads, the QD-MFS system can be directly diagonalized even
in the presence of Coulomb interaction. It finds 8 eigenstates
$\ket{q_\up,p_\down;\alpha}$ where $q_\up=0,1$ and $p_\down = e/o$ are quantum
numbers for $Q_\up$ and $P_\down$ and $\alpha = u,l$. For $\epsilon_m=0$, the
even ($e$) and odd ($o$) states are degenerate because the MFS is
energyless. $\ket{1,e/o;l}$ and $\ket{0,e/o;l}$, which would be spin-degenerate
$\ket\up$ and $\ket\down$ states at $t_m=0$, respectively, are now split [see
\figref{fig:1}(b)] with the induced Zeeman splitting
\begin{align}
  \Delta_Z
  =
  \sqrt{\frac{(\delta-\epsilon_d)^2}{4} + \frac{t_m^2}{2}}
  - \sqrt{\frac{\epsilon_d^2}{4} + \frac{t_m^2}{2}}
  - \frac{\delta}{2}
\end{align}
with $\delta\equiv 2\epsilon_d+U$. This is a genuine combined effect of
\textit{(i)} coupling to the MFS and \textit{(ii)} the Coulomb interaction, and
it is one of our key results.
Remarkably, $\Delta_Z$, having the sign opposite to $\delta$, vanishes at the
particle-hole symmetry point ($\delta=0$) since it is generated by dot charge
fluctuations in quite analogy to the exchange field induced by ferromagnetic
contacts attached to an interacting QD \cite{Choi:2004,Zitko:2012}. Upon
coupling to the leads, $\Delta_Z$ becomes renormalized to $\Delta_Z^*$, which
is larger than $\Delta_Z$: we have confirmed it by applying the Haldane's
scaling theory \cite{Haldane:1978a}. It splits the many-body resonance,
resulting in profound consequences in the Kondo state, which we will show
below.



\begin{figure}[!t]
  \centering
  \includegraphics[width=7cm]{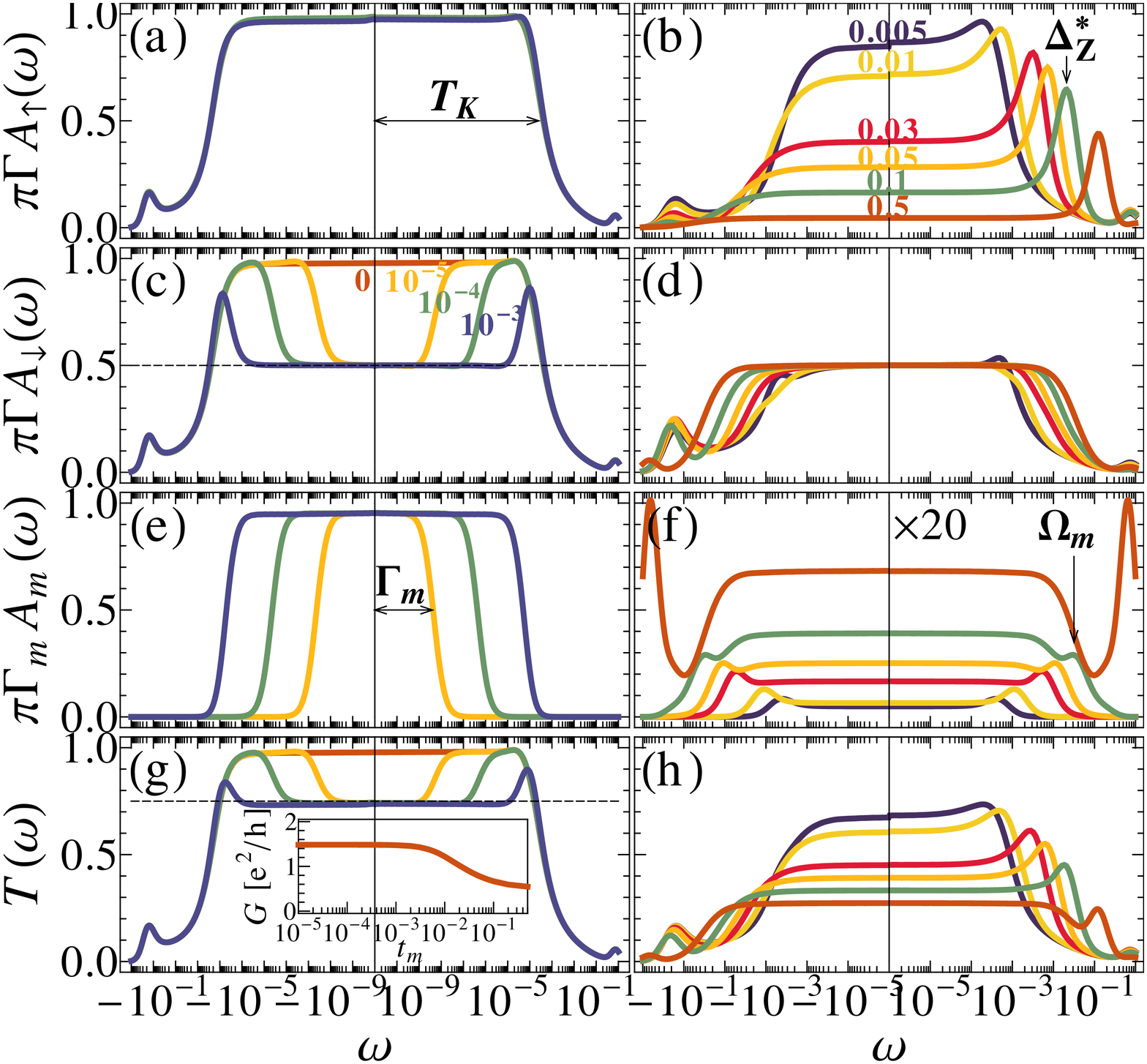}
  \caption{(Color online) Dynamical and transport features for an ideal
    dot-nanowire junction with $\epsilon_m=0$. Left and right panels represent
    the Kondo-dominant ($\Gamma_m<T_K$) and the MFS-dominant ($\Gamma_m>T_K$)
    phases, respectively: (a,b) dot spin-$\up$ spectral density, (c,d) dot
    spin-$\down$ spectral density, (e,f) $f$-operator spectral density, and
    (g,h) transmission coefficient for annotated values of $t_m$. Inset: linear
    conductance versus $t_m$. We have used $\epsilon_d=-0.2$, $U=1$,
    $\Gamma_{\rm L}=\Gamma_{\rm R}=0.02$, $D=1$, and
    $\Gamma_m=\frac{1.2\pi}{\Gamma}t_m^2$.}
  \label{fig:2}
\end{figure}

\begin{figure}[!b]
  \centering
  \includegraphics[width=4.2cm]{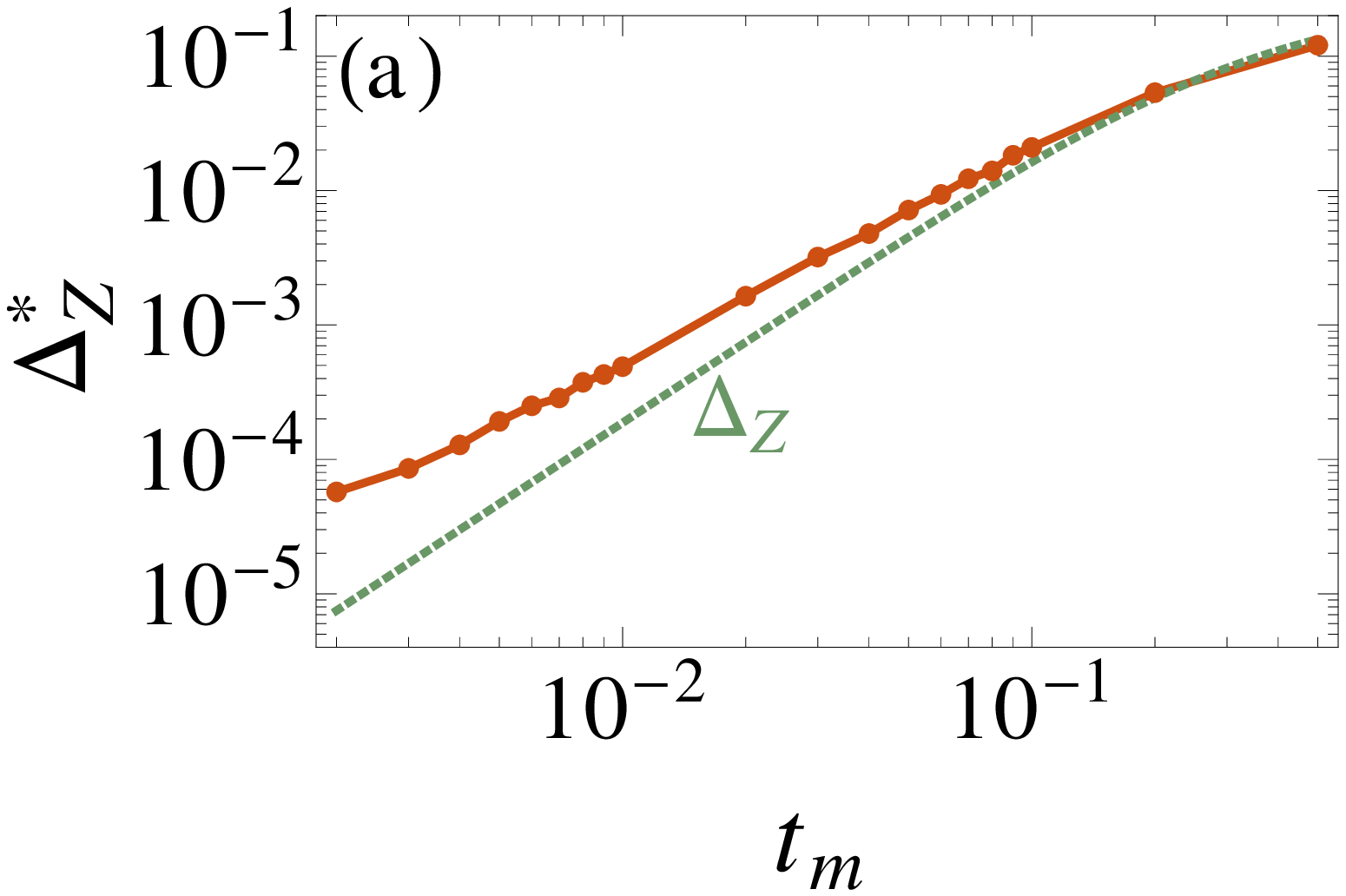}%
  \includegraphics[width=4.2cm]{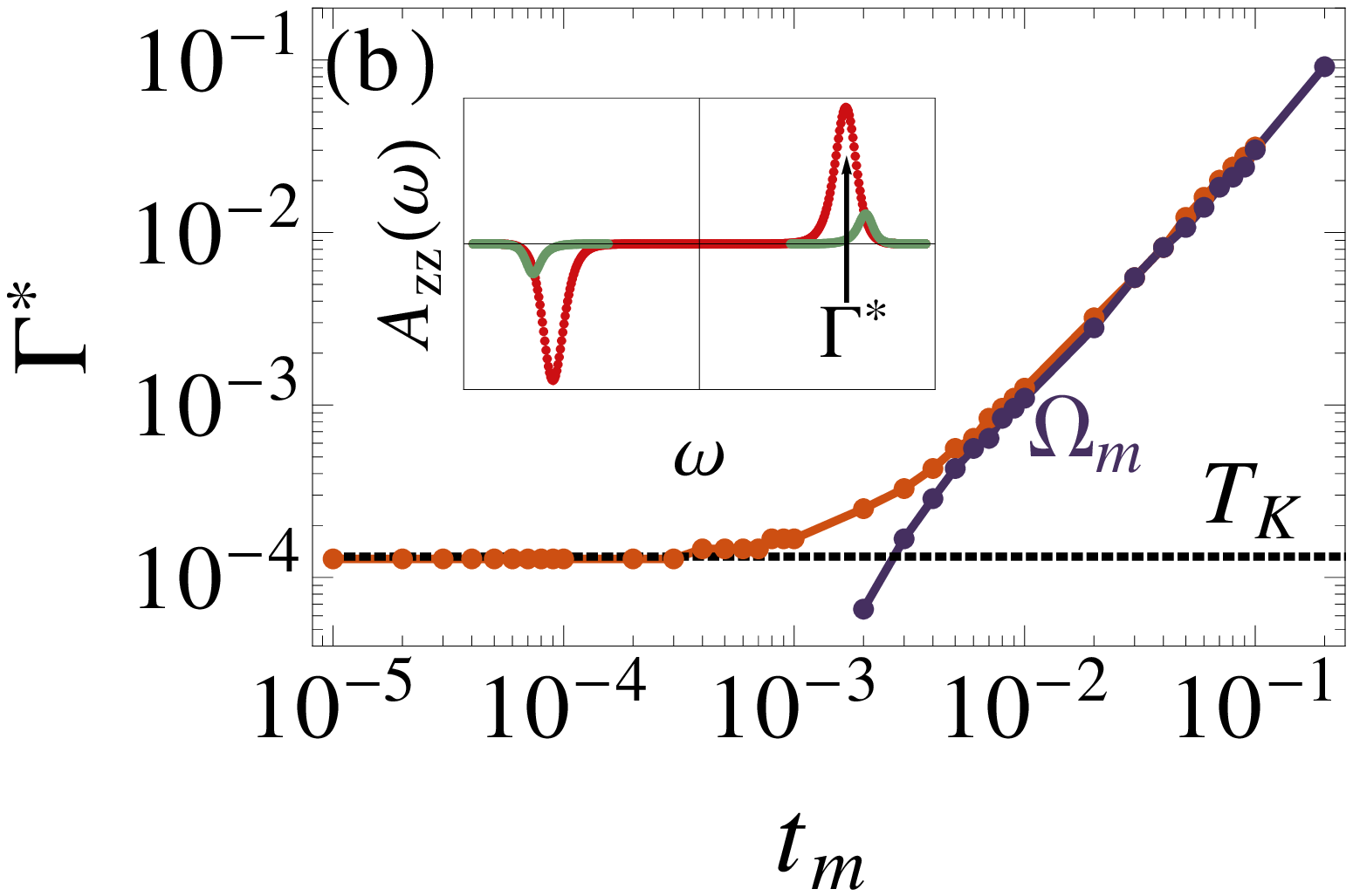}
  \caption{(Color online) (a) $\Delta_Z$ and $\Delta_Z^*$ (in the MFS-dominant
    regime) and (b) $\Gamma^*$ and $\Omega_m$ as functions of $t_m$. Inset:
    typical shape of $A_{zz}(\omega)$. We have used the same values as used in
    \figref{fig:2}.}
  \label{fig:3}
\end{figure}

\paragraph{$\epsilon_m=0$ Case.---}
We now present our NRG results for the case that the two MFSs do not overlap.
At $t_m=0$ both of $A_\mu(\omega)$ features a Kondo resonance peak centered at
$\omega = 0$ with a width $T_K$ which is the Kondo temperature:
$T_K=\sqrt{\frac{\Gamma U}{2}} \exp[\frac{\pi\epsilon_d(\epsilon_d+U)}{2\Gamma
  U}]$ \cite{Haldane:1978b}.
For $0<\Gamma_m<T_K$, where Kondo correlations are stronger than the QD-MFS
coupling, the physics resembles that of the noninteracting case, see the left
panels in \figref{fig:2}.
$A_\up(\omega)$ is intact; the Kondo peak at $\omega=0$ and two resonance peaks
at $\omega\approx\epsilon_d$, and $\omega\approx\epsilon_d+U$ are independent
of $t_m$.
The Fano-like anti-resonance due to the side-coupled MFS leads to a half-dip in
$A_\down(\omega)$, whose width is same as the width $\Gamma_m$ of the
Lorentzian-like resonance peak of $A_m(\omega)$. As in the noninteracting case
with $\epsilon_d = 0$, the value of $A_\down(\omega=0)$ is reduced by a half:
$\pi\Gamma A_\down(0)=1/2$.
Therefore, the transmission coefficients $T(\omega)$ also exhibits a Kondo peak
with a dip so that $T(\omega=0)=3/4$ and the linear conductance is pinned at
$G=3e^2/2h$.
The low-energy physics in the Kondo effect can be usually understood in terms
of a noninteracting model: a resonant level at $\epsilon^*_d = 0$ and with a
width $T_K$. The observed features above might be predictable in this
noninteracting frame. However, one should be very cautious in using the
effective theory since $t_m$ is found to be strongly renormalized. We
numerically found that $\Gamma_m = \pi\rho_{\rm dot}^* t_m^{*2} \approx
\pi \frac{c}{\Gamma} t_m^2$ where $c$ is a constant of order 1. Noting that
$\rho_{\rm dot}^* \sim 1/T_K$ in the Kondo regime, the renormalization should
lead to $t_m^* \sim \sqrt{T_K/\Gamma} t_m$ for $\Gamma_m<T_K$. The many-body
correlations are found not only to produce the Kondo effect but also to affect
the QD-MFS coupling strongly.


For $\Gamma_m>T_K$, where the QD-MFS coupling is dominant over the Kondo
effect, some peculiar features different from those in the noninteracting case
arise.
Most interestingly, the peak of $A_\up(\omega)$ shifts and get wider with
increasing $t_m$: it moves toward positive (negative) frequencies in the
electron (hole)-dominant regime, $\delta{>}0$ ($\delta{<}0$). Remarkably, in
the particle-hole symmetry point, the Kondo peak in $A_\up(\omega)$ remains at
$\omega = 0$ and only gets broader. This observation is well consistent with
the induced Zeeman splitting previously discussed. The peak position is then
identified as the renormalized Zeeman splitting $\Delta_Z^*$. We observed a
strong renormalization of $\Delta_Z$ for smaller $t_m$, see \figref{fig:3}(a):
numerically we found $\Delta_Z^*\approx 0.55 t_m^{1.5}$ for $\epsilon_d=-0.2$;
the constant factor and exponent depend on the system details.
The central peak in $A_\down(\omega)$ remains at $\omega=0$ but gets wider with
$t_m$, while $\pi\Gamma A_\down(\omega=0)$ is fixed to $1/2$. Its width is in
par with that of the central peak in $A_m(\omega)$, reflecting the coupling
between them.
Note that the side-peak structure in $A_\down(\omega)$ observed in the
noninteracting case is missing here.
Because of the shift of $A_\up(\omega)$, the transmission coefficients
$T(\omega)$ also moves its peak with $t_m$, and accordingly its value at
$\omega=0$ decreases. The linear conductance then decreases with $t_m$ for
$\Gamma_m > T_K$. However, it saturates at larger values of $t_m$ since
$A_\down(\omega=0)$ remains unchanged. In the particle-hole symmetric case
($\Delta_Z^*=0$), $A_\up$ does not move with $t_m$ so that the linear
conductance is quite independent of $t_m$ and $T(\omega)$ is fixed to $3/4$ for
wide range of frequencies centered at $\omega=0$.

\begin{figure}[!t]
  \centering
  \includegraphics[width=7cm]{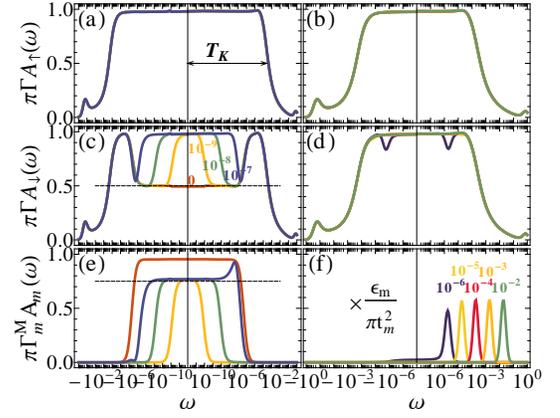}
  \caption{(Color online). Dynamical features for a realistic wire with
    $\epsilon_m\ne0$ in the Kondo dominant regime $(t_m = 10^{-4},
    \Gamma_m<T_K)$. Left and right panels represent the $\epsilon_m<\Gamma_m$
    and $\epsilon_m>\Gamma_m$ cases, respectively: (a,b) dot spin-$\up$
    spectral density, (c,d) dot spin-$\down$ spectral density, (e,f)
    $f$-operator spectral density for annotated values of $\epsilon_m$. We have
    used the same values as used in \figref{fig:2} for other parameters.}
  \label{fig:4}
\end{figure}

The spectral density $A_{zz}(\omega)$ of the spin susceptibility shows other
evidence of the impact of the MFS on the spin correlation. It has peaks at
$\omega=\pm\Gamma^*$ when the spin fluctuations are enhanced by breaking the
spin correlations, see \figref{fig:3}(b). Hence, $\Gamma^*$ should be related
to the spin binding energy. We found that $\Gamma^*=T_K$ in the Kondo dominant
regime ($\Gamma_m <T_K$), reflecting the Kondo correlation. In the MFS dominant
regime ($\Gamma_m>T_K$) we found $\Gamma^*\approx\Omega_m$, the side peak
position of $A_m(\omega)$ [see \figref{fig:2}(f)]. In the latter case, the
spin-$\down$ is more strongly hybridized with the MFS so that its correlation
energy $\Omega_m$ defines the relevant spin binding energy. Note that
$\Gamma^*$ is not directly related to $\Delta_Z^*$ which looks more relevant to
the spin correlation.  We found numerically the asymptotic relation,
$\Gamma^*\propto|t_m|^{1.45\pm0.1}$ for several values of $\epsilon_d$ in
which the exponent is rather universal.

\begin{figure}[!t]
  \centering
  \includegraphics[width=7cm]{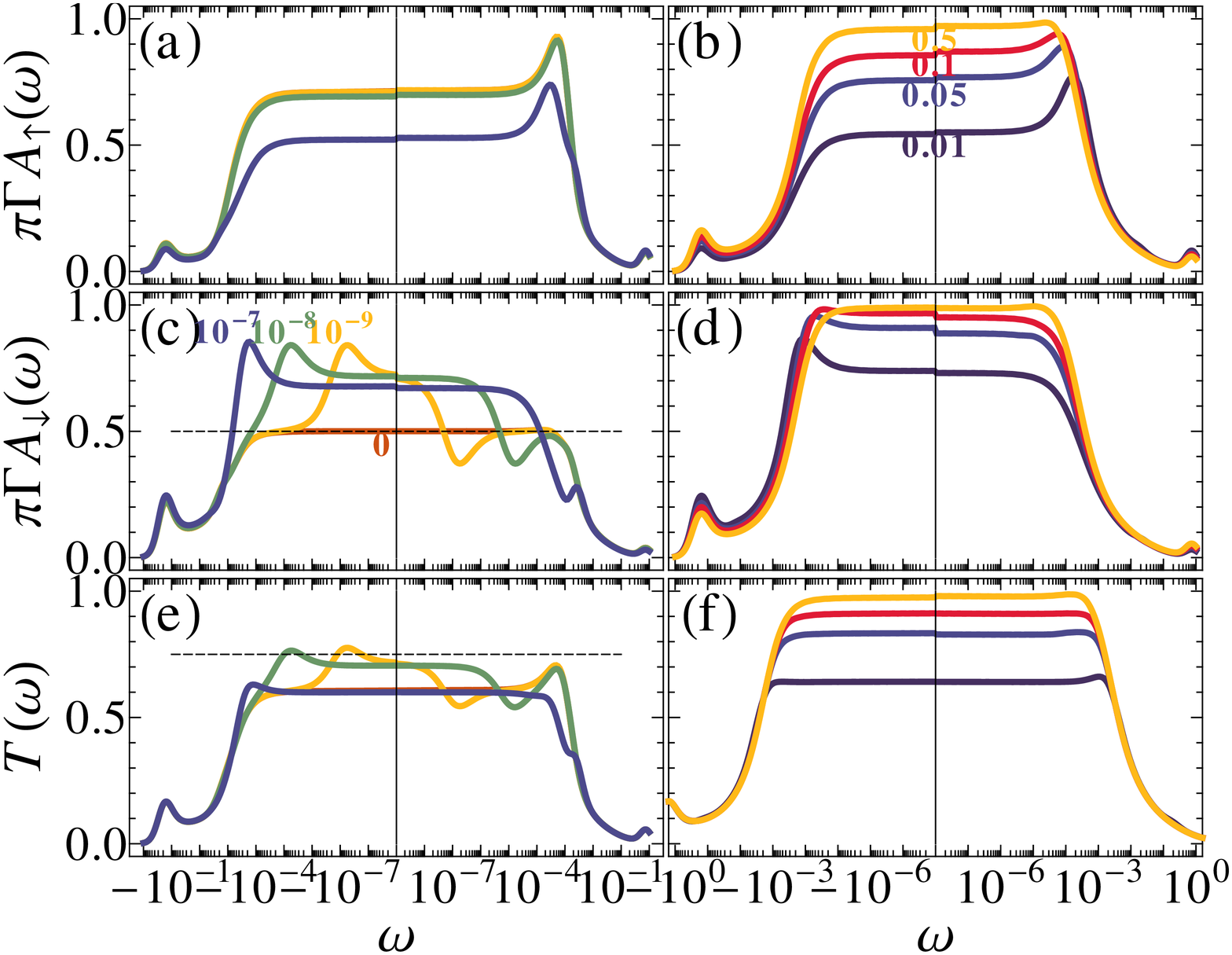}
  \caption{(Color online). Dynamical and transport features for
    $\epsilon_m\ne0$ in the MFS dominant regime $(t_m = 10^{-2}, \Gamma_m>T_K)$
    and for annotated values of $\epsilon_m$.  Left and right panels represent
    the $\epsilon_m<\Gamma_m$ and $\epsilon_m>\Gamma_m$ cases, respectively.}
  \label{fig:5}
\end{figure}

\paragraph{$\epsilon_m\neq 0$ Case. ---}
In real experiments the finite size of the wire makes the two MFSs overlap,
\emph{always} giving rise to a finite $\epsilon_m\neq 0$ or the energy
splitting between two fermionic levels $\ket0$ and $\ket1 \equiv
f^\dag\ket0$. For sufficiently large splitting, $|\epsilon_m| \gtrsim
\Gamma_m$, the fermionic levels at $\pm\epsilon_m$ does not interfere with the
Kondo resonant level formed at the Fermi level any longer so that the Kondo
physics is completely restored; see the right panels of \figsref{fig:4} and
\ref{fig:5}. Since the lower one (say $\ket0$ if $\epsilon_m>0$) is fully
occupied, only the one-way transition from $\ket0$ to $\ket1$ is possible with
respect to the $f$-fermion addition, resulting in a single-peak structure of
$A_m$ at $\omega=2\epsilon_m$ [see \figref{fig:4}(f)].

For smaller overlap ($\epsilon_m < \Gamma_m$) $A_m$ is found to retain its
Lorentzian-like peak at $\omega=0$ while its width shrinks abruptly from
$\Gamma_m$ to $\Gamma_m^M(\epsilon_m) \approx \pi\Gamma
\frac{\epsilon_m^2}{t_m^2} \approx \frac{\epsilon_m^2}{\Gamma_m}$ as soon as
$\epsilon_m$ becomes finite. It is the combined effect of strong Coulomb
interaction and the interference between two energy-split levels $\ket{0/1}$.
The splitting of $f$ levels accordingly affects the dot spectral densities.
In the Kondo dominant regime [see \figref{fig:4}(a,c,e)], while $A_\up$ is not
so affected, $A_\down$ displays a three peak structure with a central peak of a
width $\sim\Gamma_m^M$. The peak signals the disappearance of the anti-Fano
resonance by the MFS at the Fermi level. It is definitely ascribed to the shift
of the MFSs into finite energies $(\pm\epsilon_m)$. As $\epsilon_m$ grows up to
values close to $\Gamma_m$ the three peaks coalesce into a single resonance,
restoring the Kondo physics.
The linear conductance would then show an abrupt increase from $3e^2/2h$ to
$2e^2/h$ with $\epsilon_m$, which should be smoothed at finite temperatures.

In the MFS dominant regime [see \figref{fig:5}] the finite $\epsilon_m$
abolishes the half-fermionic Fano resonance at the Fermi level as wells:
$\pi\Gamma A_\down(\omega=0)$ is not pinned to $1/2$ but larger then $1/2$ for
any finite value of $\epsilon_m$. More precisely, the half-value pinning is
retained only in the finite frequencies $\Gamma_m^M < |\omega| < \Gamma_m$ for
$\epsilon_m < \Gamma_m$ [see \figref{fig:5}(c)]. Instead, the level splitting
due to the finite $\epsilon_m$ produces a peak and a small dip in the opposite
sides. The peak formed at $\omega \approx \mp\Gamma_m/\pi$ for
$\delta\gtrless0$ moves toward the higher frequencies with increasing
$\epsilon_m$ until $\epsilon_m\sim\Gamma_m$ and returns to $\omega=0$
as the Kondo physics is revived.
The effective Zeeman splitting observed in $A_\up$ also diminishes with
increasing $\epsilon_m$: the peak position gradually moves toward $\omega=0$.
Hence, in contrast to the Kondo dominant case, the restoration of the linear
conductance to the Kondo value is rather slow so that the full recovery is obtained
for $\epsilon_m \gg \Gamma_m$ [see \figref{fig:5}(f)].


\paragraph{Conclusions. ---}
We have investigated the effect of the MFS on the Kondo physics in the
side-coupled geometry. Even though the MFS is based on the superconductivity
which would suppress the Kondo effect, it is found that the Kondo effect
survives the interaction with the MFS in a modified form. We found that
coupling to the MFS introduces an electronic way to control the effective
Zeeman splitting. In addition, our results show that the energy-resolved
transmission through the dot provides an excellent way to examine the properties
of MFS and the overlap between the MFSs.

\paragraph{Acknowledgments.}
We acknowledge P. Simon and S. Andergassen for a critical reading of the manuscript.
M. L. was supported by the CAC in KHU for computer resources and the NRF grant
funded by the Korea MEST (No. 2011-0030790). R. L., and J.S. L. were supported
by MINECO Grants No. FIS2011-2352 and CSD2007-00042 (CPAN).

\end{document}